\documentclass{article}

\usepackage{color}

\definecolor{myurlcolor}{rgb}{0,0,0.5}
\usepackage{hyperref}
\hypersetup{colorlinks,
linkcolor=black,
citecolor=black,
urlcolor=myurlcolor}

\usepackage{latexsym}
\usepackage{amssymb}

\newcommand{\emptybk}{\:\:}
\newcommand{\blank}{(\emptybk)}
\newcommand{\dashbk}{-}

\newcommand{\mr}[1]{\mathrm{#1}}

\newcommand{\slsh}{/\linebreak[0]}
\newcommand{\dblslsh}{//\linebreak[0]}
\newcommand{\dt}{.\linebreak[0]}

\newcommand{\such}{\:|\:}
\newcommand{\without}{\setminus}

\newcommand{\implies}{\,\Rightarrow\,}

\def\today{\number\day\space \ifcase\month\or
  January\or February\or March\or April\or May\or June\or
  July\or August\or September\or October\or November\or December\fi
  \space\number\year}

\newcommand{\reals}{\mathbb{R}}
\newcommand{\demph}[1]{\textbf{\textup{#1}}}
\newcommand{\done}{\hfill\ensuremath{\Box}}
\newenvironment{prooflike}[1]{\begin{trivlist}\item\textbf{#1}\ }
{\end{trivlist}}
\newenvironment{proof}{\begin{prooflike}{Proof}}{\end{prooflike}}

\newcommand{\nat}{\mathbb{N}}   

\newcommand{\sub}{\subseteq}

\newcommand{\leftmat}{\left(\!\!}
\newcommand{\rightmat}{\!\!\right)}

\newlength{\stdparindent}
\newlength{\stdparskip}

\newcommand{\pv}[1]{\mathbf{#1}}
\newcommand{\transp}{\mathrm{t}}

\newcommand{\dtxt}[1]{\medskip\begin{quote}%
\setlength{\fboxsep}{2mm}%
\fbox{\parbox{0.82\textwidth}{\sl #1}}%
\end{quote}\medskip}

\newcommand{\magn}[1]{|#1|}
\newcommand{\pd}[1]{\widetilde{#1}}

\newcommand{\smp}[1]{\Delta_{#1}}
\newcommand{\ismp}[1]{\Delta_{#1}^\circ}
\newcommand{\rstr}[2]{#1|_{#2}}
\newcommand{\supp}{\mr{supp}}

\newcommand{\clique}{\mr{c}}
\newcommand{\discrete}{\mr{d}}

\newcommand{\Dmax}[1]{D_\mr{max}(#1)}
\newcommand{\ptl}[1]{\frac{\partial}{\partial #1}}

\newcommand{\passage}[1]{\paragraph{\textit{#1}}}
\newcommand{\lbl}[1]{\label{#1}}
\renewcommand{\iff}{\Leftrightarrow}
\newcommand{\bddleq}{\trianglelefteq}
\newcommand{\bddeqv}{\approx}
\newcommand{\ultraeqv}{\simeq}
\newcommand{\um}[2]{U_{#1}^{#2}}
\newcommand{\ip}[2]{\langle #1, #2 \rangle}

\newtheorem{thm}{Theorem}[section]
\newtheorem{propn}[thm]{Proposition}
\newtheorem{lemma}[thm]{Lemma}
\newtheorem{cor}[thm]{Corollary}
\newtheorem{lotsofremarks}[thm]{Remarks}

\newtheorem{predefn}[thm]{Definition}
\newenvironment{defn}{\begin{predefn}\upshape}{\end{predefn}}
\newtheorem{preexample}[thm]{Example}
\newenvironment{example}{\begin{preexample}\upshape}{\end{preexample}}
\newenvironment{example*}[1]{\begin{preexample}\upshape}{\end{preexample}}
\newtheorem{prewarning}[thm]{Warning}

\newtheorem{preexamples}[thm]{Examples}

\title{A maximum entropy theorem with applications to the measurement of
biodiversity} 
\author{Tom Leinster%
\thanks{School of Mathematics and Statistics, University of Glasgow, 
Glasgow G12 8QW, UK;
Tom.Leinster@glasgow.ac.uk.  Supported by an EPSRC Advanced Research
Fellowship.}}
\date{}

\begin{document}

\sloppy

\maketitle

\begin{abstract}
This is a preliminary article stating and proving a new maximum entropy
theorem.  The entropies that we consider can be used as measures of
biodiversity.  In that context, the question is: for a given collection
of species, which frequency distribution(s) maximize the diversity?  The
theorem provides the answer.  The chief surprise is that although we are
dealing with not just a \emph{single} entropy, but a one-parameter
\emph{family} of entropies, there is a single distribution maximizing all of
them simultaneously.
\end{abstract}

\vfill

\tableofcontents

\vfill

\dtxt{%
This article is preliminary.  Little motivation or context is given, and the
proofs are probably not optimal.  I hope to write this up in a more
explanatory and polished way in due course.}

\vfill

\newpage

\section{Statement of the problem}
\lbl{sec:statement}

\passage{Basic definitions} Fix an integer $n \geq 1$ throughout.  A
\demph{similarity matrix} is an $n\times n$ symmetric matrix $Z$ with entries
in the interval $[0, 1]$, such that $Z_{ii} = 1$ for all $i$.  A
\demph{(probability) distribution} is an $n$-tuple $\pv{p} = (p_1, \ldots,
p_n)$ with $p_i \geq 0$ for all $i$ and $\sum_{i = 1}^n p_i = 1$.

Given a similarity matrix $Z$ and a distribution $\pv{p}$, thought of as a
column vector, we may form the matrix product $Z\pv{p}$ (also a column
vector), and we denote by $(Z\pv{p})_i$ its $i$th entry.

\begin{lemma}   \lbl{lemma:positive}
Let $Z$ be a similarity matrix and $\pv{p}$ a distribution.  Then $p_i \leq
(Z\pv{p})_i \leq 1$ for all $i \in \{1, \ldots, n\}$.  In particular, if $p_i
> 0$ then $(Z\pv{p})_i > 0$.
\end{lemma}

\begin{proof}
We have
\[
(Z\pv{p})_i 
=
\sum_{j = 1}^n Z_{ij} p_j
=
p_i + \sum_{j \neq i} Z_{ij} p_j
\geq
p_i
\]
and $(Z \pv{p})_i \leq \sum_{j = 1}^n 1 p_j = 1$.
\done
\end{proof}

Let $Z$ be a similarity matrix and let $q \in [0, \infty)$.  The function
$H_q^Z$ is defined on distributions $\pv{p}$ by
\[
H^Z_q(\pv{p})
=
\left\{
\begin{array}{ll}
\displaystyle
\frac{1}{q - 1} 
\left( 1 - \sum_{i:\,p_i > 0} p_i (Z\pv{p})_i^{q - 1} \right) &
\textrm{if } q \neq 1   \\
\displaystyle
- \sum_{i:\,p_i > 0} p_i \log (Z\pv{p})_i   &
\textrm{if } q = 1.
\end{array}
\right.
\]
Lemma~\ref{lemma:positive} guarantees that these definitions are valid.  The
definition in the case $q = 1$ is explained by the fact that $H_1^Z(\pv{p}) =
\lim_{q\to 1} H_q^Z(\pv{p})$ (easily shown using l'H\^opital's rule).  We
call $H_q^Z$ the \demph{entropy of order $q$}.

\passage{Notes on the literature} The entropies $H_q^Z$ were introduced in
this generality by Ricotta and Szeidl in 2006, as an index of the diversity of
an ecological community~\cite{RS}.  Think of $n$ as the number of species,
$Z_{ij}$ as indicating the similarity of the $i$th and $j$th species, and
$p_i$ as the relative abundance of the $i$th species.  Ricotta and Szeidl used
not similarities $Z_{ij}$ but dissimilarities or `distances' $d_{ij}$; the
formulas above become equivalent to theirs on putting $Z_{ij} = 1 - d_{ij}$.

The case $Z = I$ goes back further.  Something very similar to $H_q^I$, using
logarithms to base $2$ rather than base $e$, appeared in information theory in
1967, in a paper of Havrda and Charv\'at~\cite{HC}.  Later, the entropies
$H_q^I$ were discovered in statistical ecology, in a 1982 paper of Patil and
Taillie~\cite{PT}.  Finally, they were rediscovered in physics, in a 1988
paper of Tsallis~\cite{Tsa}.

Still in the case $Z = I$, certain values of $q$ give famous quantities.  The
entropy $H_1^I$ is Shannon entropy (except that Shannon used logarithms to
base~$2$).  The entropy $H_2^I$ is known in ecology as the Simpson or
Gini--Simpson index; it is the probability that two individuals chosen at
random are of different species.

For general $Z$, the entropy of order $2$ is known as Rao's quadratic
entropy~\cite{Rao}.  It is usually stated in terms of the matrix with $(i,
j)$-entry $1 - Z_{ij}$, that is, the matrix of dissimilarities mentioned
above. 

One way to obtain a similarity matrix is to start with a finite metric space
$\{a_1, \ldots, a_n\}$ and put $Z_{ij} = e^{-d(a_i, a_j)}$.  Matrices of this
kind are investigated in depth in~\cite{MMS} and other papers cited therein.
Here, metric spaces will only appear in two examples~(\ref{eg:ultra}
and~\ref{eg:metric}).

\passage{The maximum entropy problem}  Let $Z$ be a similarity matrix and
let $q \in [0, \infty)$.  The maximum entropy problem is this:
\dtxt{%
For which distribution(s) $\pv{p}$ is
$H_q^Z(\pv{p})$ maximal, and what is the maximum value?}

The solution is given in Theorem~\ref{thm:main-entropy}.  The terms used in
the statement of the theorem will be defined shortly.  However, the following
striking fact can be stated immediately:
\dtxt{%
There is a distribution maximizing $H_q^Z$ for all
$q$ simultaneously.}
So even though the entropies of different orders rank distributions
differently, there is a distribution that is maximal for all of them.

For example, this fully explains the numerical coincidence noted in
the Results section of~\cite{AKB}.

\passage{Restatement in terms of diversity}
Let $Z$ be a similarity matrix.  For each $q \in [0, \infty)$, define a
function $D_q^Z$ on distributions $\pv{p}$ by
\begin{eqnarray*}
D_q^Z(\pv{p})   &
=       &
\left\{
\begin{array}{ll}
\left( 1 - (q - 1)H_q^Z(\pv{p}) \right)^\frac{1}{1 - q}   &
\textrm{if } q \neq 1   \\
e^{H_1^Z(\pv{p})} &
\textrm{if }q = 1
\end{array}
\right. \\
        &=      &
\left\{
\begin{array}{ll}
\displaystyle
\left(
\sum_{i:\,p_i > 0}
p_i (Z\pv{p})_i^{q - 1}
\right)^\frac{1}{1 - q}         &
\textrm{if } q \neq 1        \\
\displaystyle
\prod_{i:\,p_i > 0}
(Z\pv{p})_i^{-p_i}      &
\textrm{if } q = 1.
\end{array}
\right.
\end{eqnarray*}
We call $D_q^Z$ the \demph{diversity of order $q$}.  As for entropy, it is
easily shown that $D_1^Z(\pv{p}) = \lim_{q \to 1} D_q^Z(\pv{p})$.

These diversities were introduced informally in~\cite{EDC2}, and are explained
and developed in~\cite{MD}.  The case $Z = I$ is well known in several fields:
in information theory, $\log D_q^I$ is called the R\'enyi entropy of order
$q$~\cite{Ren}; in ecology, $D_q^I$ is called the Hill number of order
$q$~\cite{Hill}; and in economics, $1/D_q^I$ is the Hannah--Kay measure of
concentration~\cite{HK}.

The transformation between $H_q^Z$ and $D_q^Z$ is invertible and
order-preserving (increasing).  Hence the maximum entropy problem is
equivalent to the maximum diversity problem:
\dtxt{%
For which distribution(s) $\pv{p}$ is
$D_q^Z(\pv{p})$ maximal, and what is the maximum value?}
The solution is given in Theorem~\ref{thm:main}.  It will be more convenient
mathematically to work with diversity rather than entropy.  Thus, we prove
results about diversity and deduce results about entropy.

When stated in terms of diversity, a further striking aspect of the solution
becomes apparent:
\dtxt{%
There is a distribution maximizing $D_q^Z$ for all
$q$ simultaneously.  The maximum value of $D_q^Z$ is the same for all $q$.}
So every similarity matrix has an unambiguous `maximum diversity', the
maximum value of $D_q^Z$ for any $q$.

A similarity matrix may have more than one maximizing distribution---but the
collection of maximizing distributions is independent of $q > 0$.  In other
words, a distribution that maximizes $D_q^Z$ for some $q$ actually maximizes
$D_q^Z$ for all $q$ (Corollary~\ref{cor:some-all}).

The diversities $D_q^Z$ are closely related to generalized means~\cite{HLP},
also called power means.  Given a finite set $I$, positive real numbers
$(x_i)_{i \in I}$, positive real numbers $(p_i)_{i \in I}$ such that $\sum_i
p_i = 1$, and $t \in \reals$, the \demph{generalized mean} of $(x_i)_{i \in
I}$, weighted by $(p_i)_{i \in I}$, of order $t$, is
\[
\left\{
\begin{array}{ll}
\displaystyle
\left( \sum_{i \in I} p_i x_i^t \right)^{1/t}   &
\textrm{if } t \neq 0   \\
\displaystyle
\prod_{i \in I} x_i^{p_i}       &
\textrm{if } t = 0.
\end{array}
\right.
\]
For example, if $p_i = p_j$ for all $i, j \in I$ then the generalized means of
orders $1$, $0$ and $-1$ are the arithmetic, geometric and harmonic means,
respectively.  

Given a similarity matrix $Z$ and a distribution $\pv{p}$, take $I = \{ i
\in \{1, \ldots, n\} \such p_i > 0\}$.  Then $1/D_q^Z(\pv{p})$ is the
generalized mean of $((Z\pv{p})_i)_{i \in I}$, weighted by $(p_i)_{i \in I}$,
of order $q - 1$.  We deduce the following.

\begin{lemma}   \label{lemma:means}
Let $Z$ be a similarity matrix and $\pv{p}$ a distribution.  Then:
\begin{enumerate}
\item   \label{part:cts}
$D_q^Z(\pv{p})$ is continuous in $q \in [0, \infty)$
\item   \label{part:dec}
if $(Z\pv{p})_i =
(Z\pv{p})_j$ for all $i, j$ such that $p_i, p_j > 0$ then $D_q^Z(\pv{p})$ is
constant over $q \in [0, \infty)$; otherwise, $D_q^Z(\pv{p})$ is strictly
decreasing in $q \in [0, \infty)$
\item   \label{part:infty}  
$\lim_{q \to \infty} D_q^Z(\pv{p}) = 1/\max_{i: p_i > 0} (Z\pv{p})_i$.
\end{enumerate}
\end{lemma}

\begin{proof}
All of these assertions follow from standard results on generalized means.
Continuity is clear except perhaps at $q = 1$, where it follows from Theorem~3
of~\cite{HLP}.  Part~(\ref{part:dec}) follows from Theorem~16 of~\cite{HLP},
and part~(\ref{part:infty}) from Theorem~4.  
\done
\end{proof}

In the light of this, we define
\[
D_\infty^Z(\pv{p}) 
= 
1 / \max_{i:\,p_i > 0} (Z\pv{p})_i.
\]
There is no useful definition of $H_\infty^Z$, since $\lim_{q \to \infty}
H_q^Z(\pv{p}) = 0$ for all $Z$ and $\pv{p}$.

\section{Preparatory results}
\lbl{sec:prep}

Here we make some definitions and prove some lemmas in preparation for solving
the maximum diversity and entropy problems.  Some of these definitions and
lemmas can also be found in~\cite{MMS} and~\cite{AMSES}.

\emph{Convention:} for the rest of this work, unlabelled summations $\sum$ are
understood to be over all $i \in \{1, \ldots, n\}$ such that $p_i > 0$.

\passage{Weightings and magnitude}

\begin{defn}
Let $Z$ be a similarity matrix.  A \demph{weighting} on $Z$ is a column vector
$\pv{w} \in \reals^n$ such that
\[
Z\pv{w} 
= 
\leftmat
\begin{array}{c}
1       \\
\vdots  \\
1
\end{array}
\rightmat.
\]
A weighting $\pv{w}$ is \demph{non-negative} if $w_i \geq 0$ for all
$i$, and \demph{positive} if $w_i > 0$ for all $i$. 
\end{defn}

\begin{lemma}
Let $\pv{w}$ and $\pv{x}$ be weightings on $Z$.  Then $\sum_{i = 1}^n w_i =
\sum_{i = 1}^n x_i$.
\end{lemma}

\begin{proof} Write $\pv{u}$ for the column vector $(1\ \cdots\  1)^\transp$,
where $\blank^\transp$ means transpose.  Then 
\[
\sum_{i = 1}^n w_i
=
\pv{u}^\transp \pv{w}
=
(Z\pv{x})^\transp \pv{w}
=
\pv{x}^\transp (Z \pv{w})
=
\pv{x}^\transp \pv{u}
=
\sum_{i = 1}^n x_i,
\]
%
%
using symmetry of $Z$.
\done 
\end{proof}

\begin{defn}
Let $Z$ be a similarity matrix on which there exists at least one weighting.
Its \demph{magnitude} is $\magn{Z} = \sum_{i = 1}^n w_i$, for any weighting
$\pv{w}$ on $Z$. 
\end{defn}

For example, if $Z$ is invertible then there is a unique weighting $\pv{w}$ on
$Z$, and $w_i$ is the sum of the $i$th row of $Z^{-1}$.  So then
\[
\magn{Z} = \sum_{i, j = 1}^n (Z^{-1})_{ij},
\]
the sum of all $n^2$ entries of $Z^{-1}$.  This formula also appears
in~\cite{SP}, \cite{Shi} and~\cite{POP}, for closely related reasons to do
with diversity and its maximization.

\passage{Weight distributions}

\begin{defn}
Let $Z$ be a similarity matrix.  A \demph{weight distribution} for $Z$ is a
distribution $\pv{p}$ such that $(Z\pv{p})_1 = \cdots = (Z\pv{p})_n$.
\end{defn}

\begin{lemma}   \lbl{lemma:wt-distribs}
Let $Z$ be a similarity matrix.  
\begin{enumerate}
\item If $Z$ admits a non-negative weighting then $\magn{Z} > 0$.
\item If $\pv{w}$ is a non-negative weighting on $Z$ then $\pv{w}/\magn{Z}$ is
a weight distribution for $Z$, and this defines a one-to-one correspondence
between non-negative weightings and weight distributions.
\item   
If $Z$ admits a weight distribution then $Z$ admits a weighting and $\magn{Z}
> 0$.  
\item   \lbl{part:wt-distribs-reciprocal}
If $\pv{p}$ is a weight distribution for $Z$ then $(Z\pv{p})_i =
1/\magn{Z}$ for all $i$.
\end{enumerate}
\end{lemma}


\begin{proof}
\begin{enumerate}
\item Let $\pv{w}$ be a non-negative weighting.  Certainly $\magn{Z} =
\sum_{i = 1}^n w_i \geq 0$.  Since we are assuming that $n \geq 1$, the vector
$\pv{0}$ is not a weighting, so $w_i > 0$ for some $i$.  Hence $\magn{Z} > 0$.

\item The first part is clear.  To see that this defines a one-to-one
correspondence, take a weight distribution $\pv{p}$, writing $(Z\pv{p})_i = K$
for all $i$.  Since $\sum p_i = 1$, we have $p_i > 0$ for some $i$, and then
$K = (Z\pv{p})_i > 0$ by Lemma~\ref{lemma:positive}.  The vector $\pv{w} =
\pv{p}/K$ is then a non-negative weighting.

The two processes---passing from a non-negative weighting to a weight
distribution, and vice versa---are easily shown to be mutually inverse.

\item Follows from the previous parts.

\item Follows from the previous parts.
\done
\end{enumerate}
\end{proof}

The first connection between magnitude and diversity is this:
\begin{lemma}   \lbl{lemma:mag-div}
Let $Z$ be a similarity matrix and $\pv{p}$ a weight distribution for $Z$.
Then $D_q^Z(\pv{p}) = \magn{Z}$ for all $q \in [0, \infty]$.
\end{lemma}

\begin{proof}
By continuity, it is enough to prove this for $q \neq 1, \infty$.  In that
case, using Lemma~\ref{lemma:wt-distribs}(\ref{part:wt-distribs-reciprocal}),
\[
D_q^Z(\pv{p})
=
\left(
\sum p_i (Z\pv{p})_i^{q - 1}
\right)^\frac{1}{1 - q}
=
\left(
\sum p_i \magn{Z}^{1 - q}
\right)^\frac{1}{1 - q}
=
\magn{Z},
\]
as required.
\done
\end{proof}

\passage{Invariant distributions}

\begin{defn}
Let $Z$ be a similarity matrix.  A distribution $\pv{p}$ is \demph{invariant}
if $D_q^Z(\pv{p}) = D_{q'}^Z(\pv{p})$ for all $q, q' \in [0, \infty]$.
\end{defn}


Soon we will classify the invariant distributions.  To do so, we need some
more notation and a lemma.

Given a similarity matrix $Z$ and a subset $B \sub \{1, \ldots, n\}$, let
$Z_B$ be the matrix $Z$ restricted to $B$, so that $(Z_B)_{ij} = Z_{ij}$ ($i,
j \in B$).  If $B$ has $m$ elements then $Z_B$ is an $m \times m$ matrix, but
it will be more convenient to index the rows and columns of $Z_B$ by the
elements of $B$ themselves than by $1, \ldots, m$.

We will also need to consider distributions on subsets of $\{1, \ldots, n\}$.
A distribution on $B \sub \{1, \ldots, n\}$ is said to be invariant, a weight
distribution, etc., if it is invariant, a weight distribution, etc., with
respect to $Z_B$.  Similarly, we will sometimes speak of `weightings on $B$',
meaning weightings on $Z_B$.  Distributions are understood to be on $\{1,
\ldots, n\}$ unless specified otherwise.

\begin{lemma}   \lbl{lemma:absent}
Let $Z$ be a similarity matrix, let $B \sub \{1, \ldots, n\}$, and let
$\pv{r}$ be a distribution on $B$.  Write $\pv{p}$ for the distribution
obtained by extending $\pv{r}$ by zero.  Then $D_q^{Z_B}(\pv{r}) =
D_q^Z(\pv{p})$ for all $q \in [0, \infty]$.  In particular, $\pv{r}$ is
invariant if and only if $\pv{p}$ is.
\end{lemma}

\begin{proof}
For $i \in B$ we have $r_i = p_i$ and $(Z_B \pv{r})_i = (Z\pv{p})_i$.  The
result follows immediately from the definition of diversity of order $q$.
\done 
\end{proof}

By Lemma~\ref{lemma:mag-div}, any weight distribution is invariant, and by
Lemma~\ref{lemma:absent}, any extension by zero of a weight distribution is
also invariant.  We will prove that these are all the invariant distributions
there are.

For a distribution $\pv{p}$ we write $\supp(\pv{p}) = \{ i \in \{1, \ldots,
n\} \such p_i > 0\}$, the \demph{support} of $\pv{p}$.

Let $Z$ be a similarity matrix.  Given $\emptyset \neq B \sub \{1, \ldots,
n\}$ and a non-negative weighting $\pv{w}$ on $Z_B$, let $\pd{\pv{w}}$ be the
distribution obtained by first taking the weight distribution
$\pv{w}/\magn{Z_B}$ on $B$, then extending by zero to $\{1, \ldots, n\}$.

\begin{propn}   \lbl{propn:invt}
Let $Z$ be a similarity matrix and $\pv{p}$ a distribution.  The following
are equivalent:
\begin{enumerate}
\item   \lbl{part:invt-invt}
$\pv{p}$ is invariant
\item   \lbl{part:invt-support}
$(Z\pv{p})_i = (Z\pv{p})_j$ for all $i, j \in \supp(\pv{p})$
\item   \lbl{part:invt-wt-dist}
$\pv{p}$ is the extension by zero of a weight distribution on a nonempty
subset of $\{1, \ldots, n\}$  
\item   \lbl{part:invt-weighting}
$\pv{p} = \pd{\pv{w}}$ for some non-negative weighting $\pv{w}$ on some
nonempty subset of $\{1, \ldots, n\}$.
\end{enumerate}
\end{propn}

\begin{proof}
(\ref{part:invt-invt}$\implies$\ref{part:invt-support}): Follows from
Lemma~\ref{lemma:means}. 

(\ref{part:invt-support}$\implies$\ref{part:invt-wt-dist}): Suppose
that~(\ref{part:invt-support}) holds, and write $B = \supp(\pv{p})$.  The
distribution $\pv{p}$ on $\{1, \ldots, n\}$ restricts to a distribution
$\pv{r}$ on $B$.  This $\pv{r}$ is a weight distribution on
$B$, since for all $i \in B$ we have $(Z_B \pv{r})_i = (Z\pv{p})_i$,
which by~(\ref{part:invt-support}) is constant over $i \in B$.  Clearly
$\pv{p}$ is the extension by zero of $\pv{r}$.

(\ref{part:invt-wt-dist}$\implies$\ref{part:invt-invt}): Suppose that $\pv{p}$
is the extension by zero of a weight distribution $\pv{r}$ on a
nonempty subset $B \sub \{1, \ldots, n\}$.  Then for all $q \in [0, \infty]$,
\[
D_q^Z(\pv{p})
=
D_q^{Z_B}(\pv{r})
=
\magn{Z_B}
\]
by Lemmas~\ref{lemma:absent} and~\ref{lemma:mag-div} respectively; hence
$\pv{p}$ is invariant.

(\ref{part:invt-wt-dist}$\iff$\ref{part:invt-weighting}): Follows from
Lemma~\ref{lemma:wt-distribs}.
\done
\end{proof}

There is at least one invariant distribution on any given similarity matrix.
For we may choose $B$ to be a one-element subset, which has a unique
non-negative weighting $\pv{w} = (1)$, and this gives the invariant
distribution $\pd{\pv{w}} = (0, \ldots, 0, 1, 0, \ldots, 0)$.

\passage{Maximizing distributions}

\begin{defn}    \lbl{defn:max}
Let $Z$ be a similarity matrix.  Given $q \in [0,
\infty]$, a distribution $\pv{p}$ is \demph{$q$-maximizing} if $D_q^Z(\pv{p})
\geq D_q^Z(\pv{p}')$ for all distributions $\pv{p}'$.  A distribution is
\demph{maximizing} if it is $q$-maximizing for all $q \in [0, \infty]$.
\end{defn}

It makes no difference to the definition of `maximizing' if we omit $q =
\infty$; nor does it make a difference to either definition if we replace
diversity $D_q^Z$ by entropy $H_q^Z$.

We will eventually show that every similarity matrix has a
maximizing distribution.

\begin{lemma}   \lbl{lemma:max}
Let $Z$ be a similarity matrix and $\pv{p}$ an invariant distribution.  Then
$\pv{p}$ is $0$-maximizing if and only if it is maximizing.
\end{lemma}

\begin{proof}
Suppose that $\pv{p}$ is $0$-maximizing.  Then for all $q \in [0, \infty]$ and
all distributions $\pv{p}'$, 
\[
D_q^Z(\pv{p}) 
=
D_0^Z(\pv{p})
\geq
D_0^Z(\pv{p}')
\geq
D_q^Z(\pv{p}'),
\]
using invariance in the first step and Lemma~\ref{lemma:means} in the
last. 
\done
\end{proof}

\begin{lemma}   \lbl{lemma:smaller-max}
Let $Z$ be a similarity matrix and $B \sub \{1, \ldots, n\}$.  Suppose that
$\sup_{\pv{r}} D_0^{Z_B}(\pv{r}) \geq \sup_{\pv{p}} D_0^Z(\pv{p})$, where the
first supremum is over distributions $\pv{r}$ on $B$ and the second is over
distributions $\pv{p}$ on $\{1, \ldots, n\}$.  Suppose also that $Z_B$ admits
an invariant maximizing distribution.  Then so does $Z$.
\end{lemma}
(In fact, $\sup_{\pv{r}} D_0^{Z_B}(\pv{r}) \leq \sup_{\pv{p}} D_0^Z(\pv{p})$
in any case, by Lemma~\ref{lemma:absent}.  So the `$\geq$' in the statement
of the present lemma could equivalently be replaced by `$=$'.)

\begin{proof}
Let $\pv{r}$ be an invariant maximizing distribution on $Z_B$.  Define a
distribution $\pv{p}$ on $\{1, \ldots, n\}$ by extending $\pv{r}$ by zero.  By
Lemma~\ref{lemma:absent}, $\pv{p}$ is invariant.  Using
Lemma~\ref{lemma:absent} again,
\[
D_0^Z(\pv{p})
=
D_0^{Z_B}(\pv{r})
=
\sup_{\pv{r}'} D_0^{Z_B}(\pv{r}')
\geq
\sup_{\pv{p}'} D_0^Z(\pv{p}'),
\]
so $\pv{p}$ is $0$-maximizing.  Then by Lemma~\ref{lemma:max}, $\pv{p}$ is
maximizing. 
\done
\end{proof}

\passage{Decomposition}

Let $Z$ be a similarity matrix.  Subsets $B$ and $B'$ of $\{1, \ldots, n\}$
are \demph{complementary} (for $Z$) if $B \cup B' = \{1, \ldots, n\}$, $B \cap
B' = \emptyset$, and $Z_{i i'} = 0$ for all $i \in B$ and $i' \in B'$.  For
example, there exist nonempty complementary subsets if $Z$ can be expressed as
a nontrivial block sum
\[
\leftmat
\begin{array}{cc}
X       &0      \\
0       &X'
\end{array}
\rightmat.
\]

Given a distribution $\pv{p}$ and a subset $B \sub \{1, \ldots, n\}$ such that
$p_i > 0$ for some $i \in B$, let
$\rstr{\pv{p}}{B}$ be the distribution on $B$ defined by 
\[
(\rstr{\pv{p}}{B})_i 
=
\frac{p_i}{\sum_{j \in B} p_j}.
\] 

\begin{lemma}   \lbl{lemma:complementary-basic}
Let $Z$ be a similarity matrix, and let $B$ and $B'$ be nonempty complementary
subsets of $\{1, \ldots, n\}$.  Then:
\begin{enumerate}
\item   \lbl{part:complementary-weightings}
For any weightings $\pv{v}$ on $Z_B$ and $\pv{v}'$ on $Z_{B'}$,
there is a weighting $\pv{w}$ on $Z$ defined by
\[
w_i
=
\left\{
\begin{array}{ll}
v_i    &\textrm{if } i \in B  \\
v'_i   &\textrm{if } i \in B'.
\end{array}
\right. 
\]
\item   \lbl{part:complementary-invt}
For any invariant distributions $\pv{r}$ on $B$ and $\pv{r}'$ on $B'$, there
exists an invariant distribution $\pv{p}$ on $\{1, \ldots, n\}$ such that
$\rstr{\pv{p}}{B} = \pv{r}$ and $\rstr{\pv{p}}{B'} = \pv{r}'$.
\end{enumerate}
\end{lemma}

\begin{proof}
\begin{enumerate}
\item For $i \in B$, we have
\[
(Z\pv{w})_i
=
\sum_{j \in B} Z_{ij} v_j + \sum_{j \in B'} Z_{ij} v'_j
=
\sum_{j \in B} (Z_B)_{ij} v_j
=
(Z_B \pv{v})_i 
=
1.
\]
Similarly, $(Z\pv{w})_i = 1$ for all $i \in B'$.  So $\pv{w}$ is a weighting.

\item By Proposition~\ref{propn:invt}, $\pv{r} = \pd{\pv{v}}$ for some
non-negative weighting $\pv{v}$ on some nonempty subset $C \sub B$.
Similarly, $\pv{r}' = \pd{\pv{v}'}$ for some non-negative weighting $\pv{v}'$
on some nonempty $C' \sub B'$.  By~(\ref{part:complementary-weightings}),
there is a non-negative weighting $\pv{w}$ on the nonempty set $C \cup C'$
defined by 
\[
w_i
=
\left\{
\begin{array}{ll}
v_i     &\textrm{if } i \in C   \\
v'_i    &\textrm{if } i \in C'.
\end{array}
\right.
\]
Let $\pv{p} = \pd{\pv{w}}$, a distribution on $\{1, \ldots, n\}$, which is
invariant by Proposition~\ref{propn:invt}.  For $i \in C$ we have
\begin{eqnarray*}
(\rstr{\pv{p}}{B})_i    &=      &
\frac{p_i}{\sum_{j \in B} p_j}
=
\frac{w_i/\magn{Z_{C \cup C'}}}%
{\sum_{j \in C} w_j/\magn{Z_{C \cup C'}}}       \\
        &=      &
\frac{w_i}{\sum_{j \in C} w_j}
=
\frac{v_i}{\sum_{j\in C} v_j}   \\
        &=      &
r_i.
\end{eqnarray*}
For $i \in B\without C$ we have
\[
(\rstr{\pv{p}}{B})_i 
=
\frac{p_i}{\sum_{j \in B} p_j}
=
0
=
r_i.
\]
Hence $\rstr{\pv{p}}{B} = \pv{r}$, and similarly $\rstr{\pv{p}}{B'} =
\pv{r}'$.  
\done
\end{enumerate}
\end{proof}

\begin{lemma}   \lbl{lemma:complementary-D0}
Let $Z$ be a similarity matrix, let $B$
and $B'$ be complementary subsets of $\{1, \ldots, n\}$, and let $\pv{p}$ be a
distribution on $\{1, \ldots, n\}$ such that $p_i > 0$ for some $i \in B$ and
$p_i > 0$ for some $i \in B'$.  Then 
\[
D_0^Z(\pv{p})
=
D_0^{Z_B}(\rstr{\pv{p}}{B}) + 
D_0^{Z_{B'}}(\rstr{\pv{p}}{B'}).
\]
\end{lemma}

\begin{proof}
By definition,
\[
D_0^Z(\pv{p})
=
\sum \frac{p_i}{(Z\pv{p})_i}
=
\sum_{i \in B:\ p_i > 0} \frac{p_i}{(Z\pv{p})_i}
+
\sum_{i \in B':\ p_i > 0} \frac{p_i}{(Z\pv{p})_i}.
\]
Now for $i \in B$, 
\[
p_i 
=
\left( \sum_{j \in B} p_j \right)
\left( \rstr{\pv{p}}{B} \right)_i
\]
by definition of $\rstr{\pv{p}}{B}$, and 
\[
(Z\pv{p})_i 
=
\sum_{j \in B} Z_{ij} p_j + \sum_{j \in B'} Z_{ij} p_j
=
\sum_{j \in B} (Z_B)_{ij} p_j
=
\left( \sum_{j \in B} p_j \right)
(Z_B \rstr{\pv{p}}{B})_i.
\]
Similar equations hold for $B'$, so
\begin{eqnarray*}
D_0^Z(\pv{p})   &
=       &
\sum_{i \in B:\ p_i > 0} 
\frac{(\rstr{\pv{p}}{B})_i}{(Z_B \rstr{\pv{p}}{B})_i}
+
\sum_{i \in B':\ p_i > 0} 
\frac{(\rstr{\pv{p}}{B'})_i}{(Z_{B'} \rstr{\pv{p}}{B'})_i}    \\
        &=      &
D_0^{Z_B}(\rstr{\pv{p}}{B}) + 
D_0^{Z_{B'}}(\rstr{\pv{p}}{B'}),
\end{eqnarray*}
as required.
\done
\end{proof}


\begin{propn}   \lbl{propn:comp-invt-max}
Let $Z$ be a similarity matrix and let $B$ and $B'$ be nonempty complementary
subsets of $\{1, \ldots, n\}$.  Suppose that $Z_B$ and $Z_{B'}$ each admit an
invariant maximizing distribution.  Then so does $Z$.
\end{propn}

\begin{proof}
Choose invariant maximizing distributions $\pv{r}$ on $B$ and $\pv{r}'$ on
$B'$.  By
Lemma~\ref{lemma:complementary-basic}(\ref{part:complementary-invt}), there
exists an invariant distribution $\pv{p}$ on $\{1, \ldots, n\}$ such that
$\rstr{\pv{p}}{B} = \pv{r}$ and $\rstr{\pv{p}}{B'} = \pv{r}'$.  I claim that
$\pv{p}$ is maximizing.  Indeed, let $\pv{s}$ be a distribution on $\{1,
\ldots, n\}$.  If $s_i > 0$ for some $i \in B$ and $s_i > 0$ for some $i \in
B'$ then
\[
D_0^Z(\pv{s})
=
D_0^{Z_B}(\rstr{\pv{s}}{B}) + D_0^{Z_{B'}}(\rstr{\pv{s}}{B'}) 
\leq
D_0^{Z_B}(\pv{r}) + D_0^{Z_{B'}}(\pv{r}')
\]
by Lemma~\ref{lemma:complementary-D0}.  If not then without loss of
generality, $s_i = 0$ for all $i \in B'$; then $s_i > 0$ for some $i \in B$,
and
\[
D_0^Z(\pv{s})
=
D_0^{Z_B}(\rstr{\pv{s}}{B})
\leq
D_0^{Z_B}(\pv{r})
\leq 
D_0^{Z_B}(\pv{r}) + D_0^{Z_{B'}}(\pv{r}')
\]
by Lemma~\ref{lemma:absent}.  So in any case we have
\begin{eqnarray*}
D_0^Z(\pv{s})   
        &\leq   &
D_0^{Z_B}(\pv{r}) + D_0^{Z_{B'}}(\pv{r}')       \\
        &=      &
D_0^{Z_B}(\rstr{\pv{p}}{B}) + D_0^{Z_{B'}}(\rstr{\pv{p}}{B'})   \\
        &=      &
D_0^Z(\pv{p}),
\end{eqnarray*}
using Lemma~\ref{lemma:complementary-D0} in the last step.  Hence $\pv{p}$ is
$0$-maximizing, and by Lemma~\ref{lemma:max}, $\pv{p}$ is maximizing.
\done
\end{proof}

\passage{Positive definite similarity matrices}
The solution to the maximum diversity problem turns out to be simpler when the
similarity matrix is positive definite and satisfies certain further
conditions.  Here are some preparatory results.  They are not
needed for the proof of the main theorem~(\ref{thm:main}) itself, but 
will be used for the corollaries in Section~\ref{sec:cors}.

\begin{lemma}   \lbl{lemma:pos-def-basic}
Let $Z$ be a positive definite similarity matrix.  Then $Z$ has a unique
weighting and $\magn{Z} > 0$.
\end{lemma}

\begin{proof}
A positive definite matrix is invertible, so $Z$ has a unique weighting
$\pv{w}$.  By the definitions of magnitude and weighting,
\[
\magn{Z}
=
\sum_{i = 1}^n w_i
=
\pv{w}^\transp Z \pv{w}.
\]
But $n \geq 1$, so $\pv{0}$ is not a weighting, so $\pv{w} \neq \pv{0}$; then
since $Z$ is positive definite, $\pv{w}^\transp Z \pv{w} > 0$.
\done
\end{proof}


\begin{lemma}   \lbl{lemma:pos-def-sup}
Let $Z$ be a positive definite similarity matrix.  Then 
\[
\magn{Z}
=
\sup_{\pv{x}}
\frac{(\sum_{i = 1}^n x_i)^2}{\pv{x}^\transp Z \pv{x}}
\]
where the supremum is over all column vectors $\pv{x} \neq \pv{0}$.  The
points at which the supremum is attained are exactly the nonzero scalar
multiples of the unique weighting on $Z$.
\end{lemma}

\begin{proof}
Since $Z$ is positive definite, there is an inner product
$\ip{\dashbk}{\dashbk}$ on $\reals^n$ defined by
\[
\ip{\pv{x}}{\pv{y}} = \pv{x}^\transp Z \pv{y}
\]
($\pv{x}, \pv{y} \in \reals^n$).  The Cauchy--Schwarz inequality states that
for all $\pv{x}, \pv{y} \in \reals^n$,
\[
\ip{\pv{x}}{\pv{x}} \cdot \ip{\pv{y}}{\pv{y}}
\geq
\ip{\pv{x}}{\pv{y}}^2
\]
with equality if and only if one of $\pv{x}$ and $\pv{y}$ is a scalar multiple
of the other.  Let $\pv{y}$ be the unique weighting on $Z$.  Then the
inequality states that for all $\pv{x} \in \reals^n$,
\[
\pv{x}^\transp Z \pv{x} \cdot \magn{Z}
\geq
\left(\sum_{i = 1}^n x_i \right)^2.
\]
Since $\pv{y} \neq \pv{0}$, equality holds if and only if $\pv{x}$ is a scalar
multiple of $\pv{y}$.  The result follows.  
\done
\end{proof}


A vector $\pv{x}$ is \demph{nowhere zero} if $x_i \neq 0$ for all $i$.

\begin{propn}     \lbl{propn:pos-def-sub}
Let $Z$ be a positive definite similarity matrix and $B \sub \{1, \ldots,
n\}$.  Then $Z_B$ is positive definite and $\magn{Z_B} \leq \magn{Z}$.
The inequality is strict if $B$ is a proper subset and the unique weighting on
$Z$ is nowhere zero.
\end{propn}

\begin{proof}
Suppose without loss of generality that $B = \{1, \ldots, m\}$, where $0 \leq m
\leq n$.  
Let $\pv{y}$ be an $m$-dimensional column vector and write 
\[
\pv{x} 
=
(y_1, \ldots, y_m, 0, \ldots, 0)^\transp.
\]
Then
\begin{equation}        \label{eq:quad-form}
\pv{y}^\transp Z_B \pv{y}
=
\sum_{i, j = 1}^m y_i (Z_B)_{ij} y_j
=
\sum_{i, j = 1}^n x_i Z_{ij} x_j
= 
\pv{x}^\transp Z \pv{x}
\end{equation}
and
\begin{equation}        \label{eq:square}
\left( \sum_{i = 1}^m y_i \right)^2
=
\left( \sum_{i = 1}^n x_i \right)^2.
\end{equation}
By~(\ref{eq:quad-form}) and positive definiteness of $Z$, we have
$\pv{y}^\transp Z_B \pv{y} \geq 0$, with equality if and only if $\pv{x} = 0$,
if and only if $\pv{y} = 0$.  So $Z_B$ is positive definite.  Then
by~(\ref{eq:quad-form}), (\ref{eq:square}) and Lemma~\ref{lemma:pos-def-sup},
$\magn{Z_B} \leq \magn{Z}$. 

Now suppose that $m < n$ and the weighting $\pv{w}$ on $Z$ is nowhere zero.
The supremum in Lemma~\ref{lemma:pos-def-sup} is attained only at nonzero
scalar multiples of $\pv{w}$; in particular, any vector $\pv{x}$ at which it
is attained satisfies $x_n \neq 0$.  Let $\pv{y}$ be the unique weighting
on $Z_B$ and let $\pv{x}$ be the corresponding $n$-dimensional column vector,
as above.  Since $x_n = 0$, we have
\[
\magn{Z_B}
=
\frac{(\sum_{i = 1}^m y_i)^2}{\pv{y}^\transp Z_B \pv{y}}
=
\frac{(\sum_{i = 1}^n x_i)^2}{\pv{x}^\transp Z \pv{x}}
<
\magn{Z},
\]
as required.  
\done
\end{proof}

\begin{lemma}   \lbl{lemma:scattered}
Let $Z$ be a similarity matrix with $Z_{ij} < 1/(n - 1)$ for all $i \neq j$.
Then $Z$ is positive definite, and the unique weighting on $Z$ is positive.
\end{lemma}

\begin{proof}
Theorem~2 of~\cite{AMSES} shows that $Z$ is positive definite.  Now,
for each $i \in \{1, \ldots, n\}$ and $r \geq 0$, put
\[
c_{i, r}
= 
\sum_{i = i_0 \neq \cdots \neq i_r} 
Z_{i_0 i_1} Z_{i_1 i_2} \cdots Z_{i_{r - 1} i_r},
\]
where the sum is over all $i_0, \ldots, i_r \in \{1, \ldots, n\}$ such
that $i_0 = i$ and $i_{s - 1} \neq i_s$ whenever $1 \leq s \leq r$.  In
particular, $c_{i, 0} = 1$.  Write $\gamma = \max_{j \neq k} Z_{jk}$.  Then
for all $r \geq 0$,
\begin{equation}        \label{eq:wt-alt}
c_{i, r + 1} 
\leq 
\sum_{i = i_0 \neq \cdots \neq i_r \neq i_{r + 1}}
Z_{i_0 i_1} Z_{i_1 i_2} \cdots Z_{i_{r - 1} i_r} \gamma
=
(n - 1)\gamma \cdot c_{i, r}.
\end{equation}
Hence $c_{i, r} \leq ((n - 1)\gamma)^r$ for all $r \geq 0$; and $(n - 1)\gamma
< 1$, so the sum $w_i := \sum_{r = 0}^\infty (-1)^r c_{i, r}$ converges.
Again using~(\ref{eq:wt-alt}), we have $c_{i, r + 1} < c_{i, r}$ for all $r$,
so $w_i > 0$.

It remains to show that $\pv{w} = (w_1, \ldots, w_n)^\transp$ is a weighting.
Let $i \in \{1, \ldots, n\}$.  Then
\begin{eqnarray*}
(Z\pv{w})_i     &=      &
w_i + \sum_{j \neq i} Z_{ij} w_j \\
        &=      &
w_i + 
\sum_{j \neq i}  Z_{ij} 
\sum_{r = 0}^\infty (-1)^r
\sum_{j = j_0 \neq \cdots \neq j_r} Z_{j_0 j_1} \cdots Z_{j_{r-1} j_r}  \\
        &=      &
w_i - 
\sum_{r = 0}^\infty (-1)^{r + 1} c_{i, r + 1}   \\
        &=      &
w_i - (w_i - c_{i, 0})  
\\
        &=      &
1,
\end{eqnarray*}
as required.
\done
\end{proof}

\pagebreak

\section{The main theorem}

\passage{Solution to the maximum diversity problem}

\begin{thm}[Main Theorem]     \lbl{thm:main}
Let $Z$ be a similarity matrix.  Then:
\begin{enumerate}
\item   \lbl{part:main-value}
For all $q \in [0, \infty]$,
\begin{equation}        \label{eq:sup-max}
\sup_{\pv{p}} D_q^Z(\pv{p})
=
\max_B \magn{Z_B}
\end{equation}
where the supremum is over all distributions $\pv{p}$ and the
maximum is over all subsets $B \sub \{1, \ldots, n\}$ such that $Z_B$
admits a non-negative weighting.
\item   \lbl{part:main-place}
The maximizing distributions are precisely those of the form
$\pd{\pv{w}}$, where $\pv{w}$ is a non-negative weighting on a subset
$B \sub \{1, \ldots, n\}$ such that $\magn{Z_B}$ attains the
maximum~(\ref{eq:sup-max}).  
\end{enumerate}
In particular, there exists a maximizing distribution, and the maximum
diversity of order $q$ is the same for all $q \in [0, \infty]$.
\end{thm}

For the definitions, including that of `maximizing distribution', see
Section~\ref{sec:prep}.
The proof is given later in this section.  First we make some remarks on
computation and on maximum entropy. 

The \demph{maximum diversity} of a similarity matrix $Z$ is $\Dmax{Z} :=
\sup_\pv{p} D_q^Z(\pv{p})$, which by Theorem~\ref{thm:main} is independent of
the value of $q \in [0, \infty]$.

\passage{Remarks on computation}
Suppose that we are given a similarity matrix $Z$ and want to compute its
maximizing distribution(s) and maximum diversity.  The theorem gives the
following algorithm.  For each of the $2^n$ subsets $B$ of $\{1, \ldots, n\}$:
\begin{itemize}
\item perform some simple linear algebra to decide whether $Z_B$ admits a
non-negative weighting
\item if it does, tag $B$ as `good' and record the magnitude $\magn{Z_B}$
(the sum of the entries of any weighting).
\end{itemize}
The maximum of all the recorded magnitudes is the maximum
diversity $\Dmax{Z}$.  For each good $B$ such that $\magn{Z_B} =
\Dmax{Z}$, find all non-negative weightings $\pv{w}$ on $Z_B$; the
corresponding distributions $\pd{\pv{w}}$ are the maximizing distributions.

This algorithm takes exponentially many steps.  However, each step is fast, so
it might be possible to handle reasonably large values of $n$ in a reasonable
length of time.  Moreover, the results of Section~\ref{sec:cors} may allow
the speed of the algorithm to be improved.

\passage{Solution to the maximum entropy problem}
We can translate the solution to the maximum \emph{diversity} problem into a
solution to the maximum \emph{entropy} problem.  The first part, giving the
value of the maximum, becomes more complicated.  The second part, giving the
maximizing distribution(s), is unchanged.

\begin{thm}     \lbl{thm:main-entropy}
Let $Z$ be a similarity matrix.  Then:
\begin{enumerate}
\item   
For all $q \in [0, \infty)$,
\begin{equation}        \label{eq:sup-max-ent}
\sup_{\pv{p}} H_q^Z(\pv{p})
=
\left\{
\begin{array}{ll}
\max_B \frac{1}{q - 1} \left( 1 - \magn{Z_B}^{1 - q} \right)    &
\textrm{if } q \neq 1   \\
\max_B \log \magn{Z_B}  &
\textrm{if } q = 1
\end{array}
\right.
\end{equation}
where the supremum is over all distributions $\pv{p}$ and the
maxima are over all subsets $B \sub \{1, \ldots, n\}$ such that $Z_B$
admits a non-negative weighting.
\item
The maximizing distributions are precisely those of the form
$\pd{\pv{w}}$, where $\pv{w}$ is a non-negative weighting on a subset
$B \sub \{1, \ldots, n\}$ such that $\magn{Z_B}$ is maximal among all subsets
admitting a non-negative weighting.
\end{enumerate}
In particular, there exists a maximizing distribution.
\end{thm}

\begin{proof}
This follows almost immediately from Theorem~\ref{thm:main}, using the
definition of $D_q^Z$ in terms of $H_q^Z$.  Note that on the right-hand side
of~(\ref{eq:sup-max-ent}), the expressions $\frac{1}{q - 1}(1 - \magn{Z_B}^{1
- q})$ and $\log \magn{Z_B}$ are increasing, injective functions of
$\magn{Z_B}$, so a subset $B$ maximizes any one of them if and only if it
maximizes $\magn{Z_B}$.  \done
\end{proof}

The part of Theorem~\ref{thm:main} stating that the maximum diversity of order
$q$ is the same for all values of $q$ has no clean statement in terms of
entropy.

\passage{Diversity of order zero}

Our proof of Theorem~\ref{thm:main} will depend on an analysis of the function
$D_0^Z$, diversity of order zero.  The first step is to find its critical
points, and for that we need a technical lemma.

\begin{lemma}   \lbl{lemma:skew}
Let $m \geq 1$, let $Y$ be an $m \times m$ real skew-symmetric matrix, and let
$\pv{x} \in (0, \infty)^m$.  Suppose that $Y_{ij} \geq 0$ whenever $i \geq j$
and that $\sum_{j=1}^m Y_{ij} x_j$ is independent of $i \in \{1, \ldots, m\}$.
Then $Y = 0$.
\end{lemma}

\begin{proof}
This is true for $m = 1$; suppose inductively that $m \geq 2$.  We
have 
\[
\sum_{j = 1}^m Y_{1j} x_j 
= 
\sum_{j = 1}^m Y_{mj} x_j
\]
with $Y_{1j} x_j = -Y_{j1} x_j \leq 0$ and $Y_{mj} x_j \geq 0$ for all $j$;
hence both sides are $0$ and $Y_{mj} x_j = 0$ for all $j$.  So for all $j$ we
have $Y_{mj} = 0$ (since $x_j > 0$) and $Y_{jm} = 0$ (by skew-symmetry).  Let
$Y'$ be the $(m - 1) \times (m - 1)$ matrix defined by $Y'_{ij} = Y_{ij}$.
Then $Y'$ satisfies the conditions of the inductive hypothesis, so $Y' = 0$;
that is, $Y_{ij} = 0$ whenever $i, j < m$.  But we already have $Y_{ij} = 0$
whenever $i = m$ or $j = m$, so $Y = 0$, completing the induction.  \done
\end{proof}

Write
\[
\smp{n}
=
\{ \pv{p} \in \reals^n 
\such 
\sum p_i = 1,
\ p_i \geq 0 
\}
\]
for the space of distributions, and
\[
\ismp{n}
=
\{ \pv{p} \in \reals^n 
\such 
\sum p_i = 1,
\ p_i > 0 
\}
\]
for the space of nowhere-zero distributions.  The function $D_0^Z$ on
$\smp{n}$ is given by
\[
D_0^Z(\pv{p})
=
\sum \frac{p_i}{(Z\pv{p})_i}.
\]
(Recall the standing convention that unlabelled summations are over all $i \in
\{1, \ldots, n\}$ such that $p_i > 0$.)  It can be defined, using the same
formula, for all $\pv{p} \in [0, \infty)^n$.  It is then differentiable
on $(0, \infty)^n$, where the summation is over all $i \in \{1, \ldots, n\}$.

\begin{propn}
Let $Z$ be a similarity matrix and $\pv{p} \in \ismp{n}$.  Then $\pv{p}$ is a
critical point of $D_0^Z$ on $\ismp{n}$ if and only if for all $i, j \in \{1,
\ldots, n\}$,
\[
Z_{ij} > 0 \implies (Z\pv{p})_i = (Z\pv{p})_j.
\]
\end{propn}

\begin{proof}
We find the critical points of $D_0^Z$ on $\ismp{n}$ using Lagrange
multipliers and the fact that $\ismp{n}$ is the intersection of $(0,
\infty)^n$ with the hyperplane $\{ \pv{p} \in \reals^n \such \sum p_i = 1
\}$.  Write $h(\pv{p}) = \sum p_i - 1$.

For $k, i \in \{1, \ldots, n\}$ and $\pv{p} \in (0, \infty)^n$ we have
$\ptl{p_k} (Z\pv{p})_i = Z_{ik}$, giving
\[
\ptl{p_k} 
\left(
\frac{p_i}{(Z\pv{p})_i}
\right)
=
\left\{
\begin{array}{ll}
\displaystyle
\frac{1}{(Z\pv{p})_k^2} \sum_{j \neq k} Z_{kj} p_j      &
\textrm{if } k = i      \\
\displaystyle
-\frac{1}{(Z \pv{p})_i^2} Z_{ik} p_i    &
\textrm{otherwise.}
\end{array}
\right.
\]
From this and symmetry of $Z$ we deduce that for $k \in \{1, \ldots, n\}$ and
$\pv{p} \in (0, \infty)^n$,
\[
\ptl{p_k} D_0^Z(\pv{p})
=
\sum_{i = 1}^n Y_{ki} p_i
\]
where
\[
Y_{ki} 
=
\left(
\frac{1}{(Z\pv{p})_k^2} - \frac{1}{(Z\pv{p})_i^2} 
\right)
Z_{ki}.
\]
On the other hand, 
\[
\ptl{p_k} h(\pv{p})
=
1
\]
for all $k$.  A point $\pv{p} \in \ismp{n}$ is a critical point of $D_0^Z$ on
$\ismp{n}$ if and only if there exists a scalar $\lambda$ such that $(\nabla
D_0^Z)(\pv{p}) = \lambda (\nabla h)(\pv{p})$.  Hence $\pv{p}$ is critical if
and only if $\sum_i Y_{ki} p_i$ is independent of $k \in \{1, \ldots, n\}$.
So the proposition is equivalent to the statement that, for $\pv{p} \in
\ismp{n}$, the sum $\sum Y_{ki} p_i$ is independent of $k$ if and only if the
matrix $Y$ is $0$.

The `if' direction is trivial.  Conversely, suppose that $\sum_i Y_{ki} p_i$
is independent of $k$.  Assume without loss of generality that $(Z \pv{p})_1
\geq \cdots \geq (Z\pv{p})_n$.  Then Lemma~\ref{lemma:skew} applies (taking
$\pv{x} = \pv{p}$), and $Y = 0$.  \done
\end{proof}

Let $\sim$ be the equivalence relation on
$\{1, \ldots, n\}$ generated by $i \sim j$ whenever $Z_{ij} > 0$.  Thus, $i
\sim j$ if and only if there is a chain $i = i_1, i_2, \ldots, i_{r-1}, i_r =
j$ with $Z_{i_t, i_{t+1}} > 0$ for all $t$.  Call $Z$ \demph{connected} if $i
\sim j$ for all $i, j$.

\begin{cor}     \lbl{cor:conn-crit}
Let $Z$ be a connected similarity matrix.  Then every critical point of $D_0^Z$
in $\ismp{n}$ is a weight distribution.  
\done
\end{cor}

We are aiming to show, among other things, that there is a $0$-maximizing
distribution: the function $D_0^Z$ on $\smp{n}$ attains its supremum.  Its
supremum is finite, since by Lemma~\ref{lemma:positive}, $D_0^Z(\pv{p}) \leq
n$ for all distributions $\pv{p}$.  If $D_0^Z$ is continuous on $\smp{n}$ then
it certainly does attain its supremum.  But in general, it is not.  For
example, if $Z = I$ then $D_0^Z(\pv{p})$ is the cardinality of the support of
$\pv{p}$, which is not continuous in $\pv{p}$.  We must therefore use another
argument to establish the existence of a $0$-maximizing distribution.  The
following lemma will help us.

\begin{lemma}   \lbl{lemma:seqs}
Let $Z$ be a connected similarity matrix and let $(\pv{p}^{k})_{k \in \nat}$
be a sequence in $\smp{n}$.  Then $(\pv{p}^k)_{k \in \nat}$ has a subsequence
$(\pv{p}^k)_{k \in S}$ satisfying at least one of the following conditions:
\begin{enumerate}
\item   \lbl{part:seqs-boundary}
there is some $i \in \{1, \ldots, n\}$ such that $p^k_i = 0$ for all $k
\in S$
\item   \lbl{part:seqs-interior}
the subsequence $(\pv{p}^k)_{k \in S}$ lies in $\ismp{n}$ and converges
to some point of $\ismp{n}$
\item   \lbl{part:seqs-quotient}
the subsequence $(\pv{p}^k)_{k \in S}$ lies in $\ismp{n}$, and there is
some $i \in \{1, \ldots, n\}$ such that
$\lim_{k \in S} \left(p^k_i/(Z\pv{p}^k)_i\right) = 0$.
\end{enumerate}
\end{lemma}
Here and in what follows, we treat sequences as families $(x_k)_{k \in T}$
indexed over some infinite subset $T$ of $\nat$.  A subsequence of such a
sequence therefore amounts to an infinite subset of $T$.

\begin{proof}
If there exist infinitely many pairs $(k, i) \in \nat \times \{1, \ldots, n\}$
such that $p^k_i = 0$ then there is some $i \in \{1, \ldots, n\}$ such that
$\{k \in \nat \such p^k_i = 0\}$ is infinite.  Taking $S = \{k \in \nat
\such p^k_i = 0\}$ then gives condition~(\ref{part:seqs-boundary}).  

Suppose, then, that there are only finitely many such pairs.  We may choose a
subsequence $(\pv{p}^k)_{k \in Q}$ of $(\pv{p}^k)_{k \in \nat}$ lying in
$\ismp{n}$.  Further, since $\smp{n}$ is compact, we may choose a subsequence
$(\pv{p}^k)_{k \in R}$ of $(\pv{p}^k)_{k \in Q}$ converging to some point
$\pv{p} \in \smp{n}$.  If $\pv{p} \in \ismp{n}$
then $(\pv{p}^k)_{k \in R}$ satisfies~(\ref{part:seqs-interior}).

Suppose, then, that $\pv{p} \not\in \ismp{n}$; say $p_\ell = 0$ where $\ell
\in \{1, \ldots, n\}$.  

Define a binary relation $\bddleq$ on $\{1, \ldots, n\}$ by $i \bddleq j$ if
and only if $(p^k_i/p^k_j)_{k \in R}$ is bounded (that is, bounded above).
Then $\bddleq$ is reflexive and transitive, and if $i \bddleq j$ and $p_j = 0$
then $p_i = 0$.  Write $i \bddeqv j$ for $i \bddleq j \bddleq i$; then
$\bddeqv$ is an equivalence relation.

I claim that there exist $i, j \in \{1, \ldots, n\}$ with $Z_{ij} > 0$ and $i
\not\bddleq j$.  For if not, the equivalence relation $\bddeqv$ satisfies
$Z_{ij} > 0 \implies i \bddeqv j$, and since $Z$ is connected, $i \bddeqv j$
for all $i, j$.  But then $i \bddleq \ell$ for all $i$, and $p_\ell = 0$, so
$p_i = 0$ for all $i$.  This contradicts $\pv{p}$ being a distribution,
proving the claim.

Now without loss of generality, $Z_{1n} > 0$ and $1 \not\bddleq n$.  So
$(p^k_1/p^k_n)_{k \in R}$ is unbounded.  We may choose an infinite subset $S
\sub R$ such that $\lim_{k \in S} (p^k_1/p^k_n) = \infty$.  For all $k
\in S$ we have
\[
\frac{(Z\pv{p}^k)_n}{p^k_n}
\geq
\frac{Z_{n1} p^k_1}{p^k_n}
\]
with $Z_{n1} = Z_{1n} > 0$, so
\[
\lim_{k \in S}
\left(
\frac{(Z\pv{p}^k)_n}{p^k_n}
\right)
=
\infty,
\]
and condition~(\ref{part:seqs-quotient}) follows.
\done
\end{proof}

\passage{Existence of a maximizing distribution} At the heart of
Theorem~\ref{thm:main} is the following result, from which we will deduce the
theorem itself.

\begin{propn}   \lbl{propn:heart}
Every similarity matrix has a maximizing distribution, and every maximizing
distribution is invariant.  
\end{propn}

\begin{proof}
Let $Z$ be a similarity matrix.  It is enough to prove that $Z$ admits an
invariant maximizing distribution: for if $\pv{p}$ and $\pv{p}'$ are both
maximizing then $D_q^Z(\pv{p}) = D_q^Z(\pv{p}')$ for all $q$, so $\pv{p}$ is
invariant if and only if $\pv{p}'$ is.

The result holds for $n = 1$.  Suppose inductively that $n \geq 2$.

\emph{Case 1: $Z$ is not connected.} We may partition $\{1, \ldots, n\}$ into
two nonempty subsets, $B$ and $B'$, each of which is a union of
$\sim$-equivalence classes (where $\sim$ is as defined before
Corollary~\ref{cor:conn-crit}).  Then $B$ and $B'$ are complementary, and by
inductive hypothesis, $Z_B$ and $Z_{B'}$ each admit an invariant maximizing
distribution.  So by Proposition~\ref{propn:comp-invt-max}, $Z$ admits one
too. 

\emph{Case 2: $Z$ is connected.}  Write $\sigma = \sup_{\pv{p}}
D_0^Z(\pv{p})$.  We may choose a sequence $(\pv{p}^k)_{k \in \nat}$ in
$\smp{n}$ with $\lim_{k \to \infty} D_0^Z(\pv{p}^k) = \sigma$.  By
Lemma~\ref{lemma:seqs}, at least one of the following three conditions holds.
\begin{enumerate}
\item There is a subsequence $(\pv{p}^k)_{k \in S}$ such that (without loss of
generality) $p^k_n = 0$ for all $k \in S$.  Write $B = \{1, \ldots, n - 1\}$.
Define a sequence $(\pv{r}^k)_{k \in S}$ in $\smp{n - 1}$ by 
\[
\pv{r}^k = (p^k_1, \ldots, p^k_{n - 1}).
\]
Then for all $k \in S$ we have $D_0^{Z_B}(\pv{r}^k) = D_0^Z(\pv{p}^k)$ (by
Lemma~\ref{lemma:absent}), so $\sup_{k \in S} D_0^{Z_B}(\pv{r}^k) = \sigma$.
Then by Lemma~\ref{lemma:smaller-max} and inductive hypothesis, $Z$ admits an
invariant maximizing distribution.

\item There is a subsequence $(\pv{p^k})_{k \in S}$ in $\ismp{n}$ convergent
to some point $\pv{p} \in \ismp{n}$.  Since $D_0^Z$ is continuous on
$\ismp{n}$,
\[
D_0^Z(\pv{p})
=
\lim_{k \in S} D_0^Z(\pv{p}^k)
=
\sigma.
\]
So $\pv{p}$ is $0$-maximizing.  Now $\pv{p}$ is a critical point of $D_0^Z$ on
$\ismp{n}$, and $Z$ is connected, so by Corollary~\ref{cor:conn-crit},
$\pv{p}$ is a weight distribution.  By Lemma~\ref{lemma:mag-div},
$\pv{p}$ is invariant; then by Lemma~\ref{lemma:max}, $\pv{p}$ is maximizing.

\item There is a subsequence $(\pv{p}^k)_{k \in S}$ in $\ismp{n}$ such that
(without loss of generality) $\lim_{k \in S} \left( p^k_n / (Z\pv{p}^k)_n
\right) = 0$.  Write $B = \{1, \ldots, n - 1\}$.  Define a sequence
$(\pv{r}^k)_{k \in S}$ in $\ismp{n - 1}$ by $\pv{r}^k = \rstr{\pv{p}^k}{B}$
(which is possible because $\pv{p}^k \in \ismp{n}$ and $n \geq 2$).  Then for
all $k \in S$ and $i \in B$,
\[
\frac{r^k_i}{(Z_B \pv{r}^k)_i}
=
\frac{p^k_i}{\sum_{j = 1}^{n - 1} Z_{ij} p^k_j}
=
\frac{p^k_i}{(Z\pv{p}^k)_i - Z_{in} p^k_n}
\geq
\frac{p^k_i}{(Z\pv{p}^k)_i}.
\]
Hence for all $k \in S$,
\[
D_0^{Z_B}(\pv{r}^k)
=
\sum_{i = 1}^{n - 1}
\frac{r^k_i}{(Z_B \pv{r}^k)_i}
\geq
\sum_{i = 1}^{n - 1}
\frac{p^k_i}{(Z\pv{p}^k)_i}
=
D_0^Z(\pv{p}^k) - \frac{p^k_n}{(Z\pv{p}^k)_n}.
\]
But 
\[
\lim_{k \in S} 
\left(
D_0^Z(\pv{p}^k) - \frac{p^k_n}{(Z\pv{p}^k)_n}
\right)
=
\sigma - 0
=
\sigma,
\]
so $\sup_{k \in S} D_0^{Z_B}(\pv{r}^k) \geq \sigma$.   Then by
Lemma~\ref{lemma:smaller-max} and inductive hypothesis, $Z$ admits an
invariant maximizing distribution.
\end{enumerate}
So in all cases there is an invariant maximizing distribution, completing
the induction.
\done
\end{proof}

\passage{Proof of the Main Theorem,~\ref{thm:main}}
%
%
\begin{enumerate}
\item Let $q \in [0, \infty]$.  By Proposition~\ref{propn:heart}, the supremum
$\sup_\pv{p} D_q^Z(\pv{p})$ is unchanged if $\pv{p}$ is taken to run over only
the invariant distributions.  By Proposition~\ref{propn:invt}, any invariant
distribution is of the form $\pd{\pv{w}}$ for some non-negative weighting
$\pv{w}$ on some nonempty subset $B \sub \{1, \ldots, n\}$.  Hence 
\[
\sup_{\pv{p}} D_q^Z(\pv{p})
=
\max_{B, \pv{w}} D_q^Z(\pd{\pv{w}})
\]
where the maximum is over all nonempty $B$ and non-negative weightings
$\pv{w}$ on $B$.  But for any such $B$ and $\pv{w}$ we have 
\[
D_q^Z(\pd{\pv{w}}) 
=  
D_q^{Z_B}(\pv{w}/\magn{Z_B})
=
\magn{Z_B}
\]
by Lemmas~\ref{lemma:absent} and~\ref{lemma:mag-div} respectively.  Hence 
\[
\sup_{\pv{p}} D_q^Z(\pv{p})
=
\max_B \magn{Z_B}
\]
where the maximum is now over all nonempty $B \sub \{1, \ldots, n\}$ such that
there exists a non-negative weighting on $Z_B$.  And since $\magn{\emptyset} =
0$, it makes no difference if we allow $B$ to be empty.

\item Any maximizing distribution is invariant, by
Proposition~\ref{propn:heart}.  The result now follows from
Proposition~\ref{propn:invt}.  
\done
\end{enumerate}

\section{Corollaries and examples}
\lbl{sec:cors}

Here we state some corollaries to the results of the previous section.  The
first is a companion to Lemma~\ref{lemma:max}.

\begin{cor}     \lbl{cor:some-all}
Let $Z$ be a similarity matrix and $q \in (0, \infty]$.  Then a distribution
is $q$-maximizing if and only if it is maximizing.
\end{cor}

In other words, if a distribution is $q$-maximizing for \emph{some}
$q > 0$ then it is $q$-maximizing for \emph{all} $q \geq 0$.  The proof is
below. 

However, a $0$-maximizing distribution is not necessarily maximizing.  Take $Z
= I$, for example.  Then $D_0^Z(\pv{p}) = \sum_{i:\,p_i > 0} 1 = $ cardinality
of $\supp(\pv{p})$, so any nowhere-zero distribution is $0$-maximizing.  On
the other hand, only the uniform distribution $(1/n, \ldots, 1/n)$ is
maximizing.  So the restriction $q \neq 0$ cannot be dropped from
Corollary~\ref{cor:some-all}, nor can the word `invariant' be dropped
from Lemma~\ref{lemma:max}.

\begin{proof}
Let $\pv{p}$ be a $q$-maximizing distribution.  Then
\[
D_q^Z(\pv{p})
=
\Dmax{Z}
\geq
D_0^Z(\pv{p})
\geq
D_q^Z(\pv{p}),
\]
where the second inequality is by Lemma~\ref{lemma:means}.  So we have
equality throughout, and in particular $D_0^Z(\pv{p}) = D_q^Z(\pv{p})$.  But
$q > 0$, so by Lemma~\ref{lemma:means}, $\pv{p}$ is invariant.  Hence for all
$q' \in [0, \infty]$,
\[
D_{q'}^Z(\pv{p}) = D_q^Z(\pv{p}) = \Dmax{Z},
\]
and therefore $\pv{p}$ is maximizing.
\done 
\end{proof}

The importance of Corollary~\ref{cor:some-all} is that if one has solved the
problem of maximizing entropy or diversity of any particular order $q > 0$,
then one has solved the problem of maximizing entropy and diversity of all
orders.  In the following example, we observe that for a certain class of
similarity matrices, the problem of maximizing the entropy of order $2$ has
already been solved in the literature; we can immediately deduce a
more general maximum entropy theorem.

\begin{example} \lbl{eg:graphs}
Let $G$ be a finite reflexive graph.  Thus, $G$ consists of a finite set $\{1,
\ldots, n\}$ of vertices equipped with a reflexive symmetric binary relation
$E$.  Such graphs correspond to similarity matrices $Z$ whose entries are all
$0$ or $1$, taking $Z_{ij} = 1$ if $(i, j) \in E$ and $Z_{ij} = 0$
otherwise.

For each $q \in [0, \infty]$ and distribution $\pv{p}$, define
$D_q^G(\pv{p}) = D_q^Z(\pv{p})$, the \demph{diversity of order $q$} of
$\pv{p}$ with respect to $G$.  Thus,
\[
D_q^G(\pv{p})
=
\left\{
\renewcommand{\arraystretch}{3}
\begin{array}{ll}
\displaystyle
\left( \sum_{i:\,p_i > 0} p_i
\left( \sum_{j:\,(i, j) \in E} p_j \right)^{q - 1}
\right)^\frac{1}{1 - q} &
\textrm{if } q \neq 1, \infty   \\
\displaystyle
\prod_{i:\,p_i > 0} 
\left( \sum_{j:\,(i, j) \in E} p_j \right)^{-p_i}        &
\textrm{if } q = 1      \\
\displaystyle
1/\max_{i:\,p_i > 0} \sum_{j:\,(i, j) \in E} p_j          &
\textrm{if } q = \infty.
\end{array}
\renewcommand{\arraystretch}{1}
\right.
\]

A set $K \sub \{1, \ldots, n\}$ of vertices of $G$ is \demph{discrete} if $(i,
j) \not\in E$ whenever $i, j \in K$ with $i \neq j$.  Write $\discrete(G)$
for the largest integer $d$ such that there exists a discrete set in $G$
of cardinality $d$.  Also, given any
nonempty set $B \sub \{1, \ldots, n\}$, write $\pv{p}^B$ for the distribution
\[
\pv{p}^B_i
=
\left\{
\begin{array}{ll}
1/|B|   &\textrm{if } i \in B   \\
0       &\textrm{otherwise}.
\end{array}
\right.
\]

\emph{Claim:} For all $q \in [0, \infty]$,
\[
\sup_{\pv{p}} D_q^G(\pv{p}) = \discrete(G),
\]
and the supremum is attained at $\pv{p}^K$ for any discrete set $K$ of
cardinality $\discrete(G)$.

\emph{Proof:} We use the following result of Berarducci, Majer and
Novaga~\cite{BMN}.  Let $G'$ be a finite \emph{ir}reflexive graph with $n$
vertices, that is, an irreflexive symmetric binary relation $E'$ on $\{1,
\ldots, n\}$.  A set $K$ of vertices of $G'$ is a \demph{clique} (or complete
subgraph) if $(i, j) \in E'$ whenever $i, j \in K$ with $i \neq j$.  Write
$\clique(G')$ for the largest integer $c$ such that there exists a clique in
$G'$ of cardinality $c$.  Their Proposition~4.1 states that
\[
\sup_{\pv{p}} \sum_{(i, j) \in E'} p_i p_j
=
1 - \frac{1}{\clique(G')}
\]
(which they call the `capacity' of $G'$).  Their proof shows that the supremum
is attained at $\pv{p}^K$ for any clique $K$ of cardinality
$\clique(G')$. 

We are given a graph $G$.  Let $G'$ be its dual graph, with the same vertex-set
and with edge-relation $E'$ defined by $(i, j) \in E'$ if and only if $(i, j)
\not\in E$.  Then $G'$ is irreflexive, a clique in $G'$ is the same as a
discrete set in $G$, and $\clique(G') = \discrete(G)$.  For any distribution
$\pv{p}$,
\[
\sum_{(i, j) \in E'} p_i p_j 
=
1 - \sum_{(i, j) \in E} p_i p_j
=
H_2^Z(\pv{p}).
\]
Let $K$ be a clique in $G'$ of maximal cardinality, that is, a discrete set in
$G$ of maximal cardinality.  Then by~\cite{BMN}, $\pv{p}^K$ is 2-maximizing and
\[
H_2^Z(\pv{p}^K) 
= 
1 - \frac{1}{\clique(G')} 
=
1 - \frac{1}{\discrete(G)}.
\]
But it is a completely general fact that 
\[
H_2^Z(\pv{p}) 
= 
1 - \frac{1}{D_2^Z(\pv{p})}
\]
for all $\pv{p}$, directly from the definitions in
Section~\ref{sec:statement}.  Hence $\discrete(G) = D_2^Z(\pv{p}^K) =
D_2^G(\pv{p}^K)$, and the claim holds for $q = 2$.
Corollary~\ref{cor:some-all} now tells us that the claim holds for all $q \in
[0, \infty]$.

This class of examples tells us that a similarity matrix may have several
different maximizing distributions, and that a maximizing distribution
$\pv{p}$ may have $p_i = 0$ for some values of $i$.  These phenomena have
been observed in the ecological literature in the case $q = 2$ (Rao's
quadratic entropy): see Pavoine and Bonsall~\cite{PB} and references therein. 
\end{example}

Computing the maximum diversity is potentially slow, because in principle
one has to go through all $2^n$ subsets of $\{1, \ldots, n\}$.  But if the
similarity matrix satisfies some further conditions, a maximizing distribution
can be found very quickly:

\begin{cor}     \lbl{cor:pos-def-div}
Let $Z$ be a positive definite similarity matrix whose unique weighting
$\pv{w}$ is non-negative.  Then $\Dmax{Z} = \magn{Z}$.  Moreover,
$\pv{w}/\magn{Z}$ is a maximizing distribution, and if $\pv{w}$ is
positive then it is the unique such.
\end{cor}

\begin{proof}
Follows immediately from Proposition~\ref{propn:pos-def-sub} and
Theorem~\ref{thm:main}. 
\done
\end{proof}

\begin{example} \lbl{eg:ultra}
A similarity matrix $Z$ is \demph{ultrametric} if $\min\{Z_{ij}, Z_{jk}\} \leq
Z_{ik}$ for all $i, j, k$ and $Z_{ij} < 1$ for all $i \neq j$.  As shown
below, every ultrametric matrix is positive definite and its weighting
is positive.  Hence its maximum diversity is its magnitude, it has a unique
maximizing distribution, and that distribution is nowhere zero.

Ultrametric matrices are closely related to \demph{ultrametric spaces}, that
is, metric spaces satisfying a stronger version of the triangle inequality:
\[
\max\{ d(a, b), d(b, c) \} \geq d(a, c)
\]
for all points $a, b, c$.  Any finite metric space $A = \{a_1, \ldots, a_n\}$
gives rise to a similarity matrix $Z$ by putting $Z_{ij} = e^{-d(a_i, a_j)}$,
and if the space $A$ is ultrametric then so is the matrix $Z$.

Ultrametric matrices also arise in the quantification of biodiversity.  Take a
collection of $n$ species, and suppose, for example, that we choose a
taxonomic measure of species similarity:
\[
Z_{ij} =
\left\{
\begin{array}{ll}
1       &\textrm{if } i = j     \\
0.8     &\textrm{if $i \neq j$ but the $i$th and $j$th species are of the same
genus} \\
0.6     &\textrm{if the $i$th and $j$th species are of different genera but
the same family}        \\
0       &\textrm{otherwise.}
\end{array}
\right.
\]
This is an ultrametric matrix, so is guaranteed to have a unique maximizing
distribution.  That distribution is nowhere zero: maximizing diversity
does not eradicate any species.  The same conclusion for general ultrametric
matrices was reached, in the case $q = 2$, by Pavoine, Ollier and
Pontier~\cite{POP}. 

We now prove that any ultrametric matrix $Z$ is positive definite with
positive weighting.  That $Z$ is positive definite was also proved by Varga
and Nabben~\cite{VN}, and that the weighting is positive was also proved
in~\cite{POP}.  The following proof, which is probably not new either, seems
more direct.

If $n = 1$ then certainly $Z$ is positive definite and its weighting is
positive.  

Suppose inductively that $n \geq 2$.  Write $z = \min_{i, j} Z_{ij} < 1$.  By
the ultrametric property, there is an equivalence relation $\ultraeqv$ on
$\{1, \ldots, n\}$ defined by $i \ultraeqv j$ if and only if $Z_{ij} > z$.  We
may partition $\{1, \ldots, n\}$ into two nonempty subsets, $B$ and $B'$, each
of which is a union of $\ultraeqv$-equivalence classes; and without loss of
generality, $B = \{1, \ldots, m\}$ and $B' = \{m + 1, \ldots, n\}$, where $1
\leq m < n$.  For all $i \leq m$ and $j \geq m + 1$ we have $Z_{ij} \leq z$,
that is, $Z_{ij} = z$.  Hence 
\[
Z 
=
\leftmat
\begin{array}{cc}
Y               &z\um{m}{n - m} \\
z\um{n - m}{m}  &Y'
\end{array}
\rightmat
\]
where $Y$ is some $m \times m$ matrix, $Y'$ is some $(n - m) \times (n - m)$
matrix, and $\um{k}{\ell}$ denotes the $k \times \ell$ matrix all of whose
entries are $1$.  Now $Y$ and $Y'$ are ultrametric with entries in $[z, 1]$,
so the matrices
\[
X = \frac{1}{1 - z} (Y - z\um{m}{m}),
\qquad
X' = \frac{1}{1 - z}(Y' - z\um{n - m}{n - m})
\]
are also ultrametric.  By inductive hypothesis, $X$ and $X'$ are
positive definite and their respective weightings are positive.

We have
\begin{equation}        \lbl{eq:ultra-inductive}
Z  
= 
z \um{n}{n} + 
(1 - z)
\leftmat
\begin{array}{cc}
X       &0      \\
0       &X'
\end{array}
\rightmat.
\end{equation}
The matrix $\um{n}{n}$ is positive-semidefinite, since $\pv{x}^\transp
\um{n}{n} \pv{x} = (x_1 + \cdots + x_n)^2$ for all $\pv{x} \in \reals^n$.
Also $\leftmat
\begin{array}{cc}
X       &0      \\
0       &X'
\end{array}
\rightmat$ is positive definite, since $X$ and $X'$ are.  Finally, $z \geq 0$
and $1 - z > 0$.  It follows that $Z$ is positive definite.

Write $\pv{v}$ and $\pv{v}'$ for the weightings on $X$ and $X'$.  Put
\[
\pv{w}
=
\frac{1}{z\left(\sum_{i = 1}^m v_i + \sum_{j = 1}^{n - m} v'_j\right) 
+ (1 - z)}
\leftmat
\begin{array}{c}
v_1     \\ 
\vdots  \\
v_m     \\
v'_1    \\
\vdots  \\
v'_{n - m}
\end{array}
\rightmat.
\]
The weightings $\pv{v}$ and $\pv{v'}$ are positive and $0 \leq z \leq 1$, so
$\pv{w}$ is positive.  And it is routine to verify,
using~(\ref{eq:ultra-inductive}), that $\pv{w}$ is the weighting on $Z$.
\end{example}

\begin{example} \lbl{eg:metric}
Take a metric space with three points, $a_1, a_2, a_3$, and put $Z_{ij}
= e^{-d(a_i, a_j)}$.  This defines a $3 \times 3$ similarity matrix $Z$ with
$Z_{ij} < 1$ for all $i \neq j$ and $Z_{ij} Z_{jk} \leq Z_{ik}$ for all $i, j,
k$.  We will show that $Z$ is positive definite and that its unique weighting
is positive.  It follows that there is a unique maximizing distribution and
that the maximum diversity is $\magn{Z}$.  We give explicit expressions for
both.

First, Sylvester's Criterion states that a symmetric real $n\times n$ matrix
is positive definite if and only if for all $m \in \{1, \ldots, n\}$, the
upper-left $m \times m$ submatrix has positive determinant.  In this case:
\begin{itemize}
\item the upper-left $1 \times 1$ matrix is $(1)$, which has determinant $1$
\item the upper-left $2 \times 2$ matrix is $\leftmat \begin{array}{cc} 1
&Z_{12}\\ Z_{12} &1 \end{array} \rightmat$, which has determinant $1 -
Z_{12}^2 > 0$
\item the upper-left $3 \times 3$ matrix is $Z$ itself, and
\begin{eqnarray*}
\det Z  &=      &
1 - (Z_{12}^2 + Z_{23}^2 + Z_{31}^2) + 2 Z_{12} Z_{23} Z_{31}   \\
        &=      &
(1 - Z_{12})(1 - Z_{23})(1 - Z_{31}) 
+
(1 - Z_{12})(Z_{12} - Z_{13} Z_{32})    \\
        &       &
{}+
(1 - Z_{23})(Z_{23} - Z_{21} Z_{13})
+
(1 - Z_{31})(Z_{31} - Z_{32} Z_{21})    \\
        &>      &
0.
\end{eqnarray*}
\end{itemize}
Hence $Z$ is positive definite.  Next, it is easily checked that the
unique weighting $\pv{w}$ is given by $\pv{w} = \pv{v}/\det Z$, where, for
instance, 
\begin{eqnarray*}
v_1     &=      &
1 - (Z_{12} + Z_{13}) + (Z_{13}Z_{32} + Z_{12}Z_{23}) - Z_{23}^2    \\
        &=      &
(1 - Z_{12})(1 - Z_{23})(1 - Z_{31})
+
(1 - Z_{23})(Z_{23} - Z_{21}Z_{13})     \\
        &>      &
0.
\end{eqnarray*}
Since $\det Z > 0$, the weighting $\pv{w}$ is positive.  

The maximum diversity is $\magn{Z} = w_1 + w_2 + w_3$, which is
\[
1 +
\frac{2(1 - Z_{12})(1 - Z_{23})(1 - Z_{31})}%
{1 - (Z_{12}^2 + Z_{23}^2 + Z_{31}^2) + 2 Z_{12}Z_{23}Z_{31}}.
\]
(This expression was pointed out to me by Simon Willerton.)  The unique
maximizing distribution $\pv{p}$ is given by $\pv{p} = \pd{\pv{w}} =
\pv{w}/\magn{Z} = \pv{v}/(\magn{Z}\det Z)$, so
\[
p_1
=
\frac{1 - (Z_{12} + Z_{13}) + (Z_{13}Z_{32} + Z_{12}Z_{23}) - Z_{23}^2}%
{1 - (Z_{12}^2 + Z_{23}^2 + Z_{31}^2) + 2 Z_{12}Z_{23}Z_{31}
+ 2(1 - Z_{12})(1 - Z_{23})(1 - Z_{31})}
\]
and similarly for $p_2$ and $p_3$.
\end{example}

Example~\ref{eg:graphs} (graphs) shows that maximizing distributions
sometimes contain some zero entries.  In ecological terms this means that
diversity is sometimes maximized by completely eradicating certain species,
which may be contrary to acceptable practice.  For this and other reasons, we
might seek conditions under which some or all of the maximizing distributions
$\pv{p}$ satisfy $p_i > 0$ for all $i$.

\begin{cor}     \lbl{cor:close-to-id}
Let $Z$ be a similarity matrix such that $Z_{ij} < 1/(n - 1)$ for all $i \neq
j$.  Then $\Dmax{Z} = \magn{Z}$.  Moreover, $Z$ has a unique weighting
$\pv{w}$, the unique maximizing distribution is $\pv{p} = \pv{w}/\magn{Z}$,
and $p_i > 0$ for all $i$.
\end{cor}

\begin{proof}
By Lemma~\ref{lemma:scattered}, $Z$ is positive definite and its unique
weighting is positive.  Then apply Corollary~\ref{cor:pos-def-div}.
\done
\end{proof}

The extra hypothesis on $Z$ is strong, possibly too strong for the corollary
to be of any use in ecology: when $n$ is large, it forces $Z$ to be very close
to the identity matrix.  On the other hand, the ecological interpretation of
Corollary~\ref{cor:close-to-id} is clear: if we treat every species as highly
dissimilar to every other, the distribution that
maximizes diversity conserves all of them.

\passage{Acknowledgements} Parts of this work were done during visits to the
Centre de Recerca Matem\`atica, Barcelona, and the School of Mathematics and
Statistics at the University of Sheffield.  I am grateful to both institutions
for their hospitality, and to Eugenia Cheng, David Jordan and Joachim Kock for
making my visits both possible and pleasant.  I thank Christina Cobbold,
Andr\'e Joyal and Simon Willerton for useful conversations, and Ji\v{r}\'\i\
Velebil for tracking down a reference.


\begin{thebibliography}{MMM}


\bibitem[AKB]{AKB}
B. Allen, M. Kon, Y. Bar-Yam,
A new phylogenetic diversity measure generalizing the Shannon index and its
application to phyllostomid bats, 
\emph{The American Naturalist} 174 (2009), 236--243.

\bibitem[BMN]{BMN}
A. Berarducci, P. Majer, M. Novaga,
Percolation-type problems on infinite random graphs,
\href{http://arxiv.org/abs/0809.2335}{arXiv:0809.2335} (2008), available from
\href{http://arxiv.org}{http:\dblslsh arxiv\dt org}.

\bibitem[HK]{HK}
L. Hannah, J. Kay,
\emph{Concentration in the Modern Industry: Theory, Measurement, and the U.K.
Experience},
MacMillan, 1977.

\bibitem[HLP]{HLP}
G.H. Hardy, J.E. Littlewood, G. P\'olya,
\emph{Inequalities},
second edition,
Cambridge University Press, 1952. 

\bibitem[HC]{HC}
J. Havrda, F. Charv\'at,
Quantification method of classification processes,
\emph{Kybernetika} 3 (1967), 30--35. 

\bibitem[Hill]{Hill}
M.O. Hill,
Diversity and evenness: a unifying notation and its consequences,
\emph{Ecology}
54 (1973), No.\ 2, 427--432.


\bibitem[Lei1]{EDC2}
T. Leinster,
Entropy, diversity and cardinality (part 2),
post at \emph{The $n$-Category Caf\'e}, 7 November 2008,  
\href{http://golem.ph.utexas.edu/category/2008/11/entropy_diversity_and_cardinal_1.html}{http:\dblslsh
golem\dt ph\dt utexas\dt edu\slsh category\slsh 2008\slsh 11\slsh
entropy\_diversity\_and\_cardinal\_1\dt html}

\bibitem[Lei2]{MMS}
T. Leinster,
The magnitude of metric spaces,
\href{http://arxiv.org/abs/1012.5857}{arXiv:1012.5857} (2010), available from
\href{http://arxiv.org}{http:\dblslsh arxiv\dt org}.

\bibitem[LC]{MD}
T. Leinster, C.A. Cobbold,
Measuring diversity: the importance of species similarity,
submitted, 2010.

\bibitem[LW]{AMSES} 
T. Leinster, S. Willerton,
On the asymptotic magnitude of subsets of Euclidean space,
\href{http://arxiv.org/abs/0908.1582}{arXiv:0908.1582} (2009), available from
\href{http://arxiv.org}{http:\dblslsh arxiv\dt org}.

\bibitem[PT]{PT}
G.P. Patil, C. Taillie,
Diversity as a concept and its measurement,
\emph{Journal of the American Statistical Association} 
77 (1982), No.\ 379, 548--561. 

\bibitem[PB]{PB}
S. Pavoine, M.B. Bonsall,
Biological diversity: distinct distributions can lead to the maximization of
Rao's quadratic entropy,
\emph{Theoretical Population Biology}
75 (2009), 153--163.

\bibitem[POP]{POP}
S. Pavoine, S. Ollier, D. Pontier, 
Measuring diversity from dissimilarities with Rao's quadratic entropy: are any
dissimilarities suitable?
\emph{Theoretical Population Biology}
67 (2005), 231--239.

\bibitem[Rao]{Rao}
C.R. Rao,
Diversity and dissimilarity coefficients: a unified approach,
\emph{Theoretical Population Biology} 
21 (1982), 24--43.

\bibitem[R\'en]{Ren}
A. R\'enyi,
On measures of entropy and information,
in J. Neyman (ed.), 
\emph{Proceedings of the 4th Berkeley Conference on Mathematical Statistics
and Probability}, 
pages 547--561,
University of California Press, 1961.

\bibitem[RS]{RS}
C. Ricotta, L. Szeidl,
Towards a unifying approach to diversity measures: bridging the gap between
the Shannon entropy and Rao's quadratic index,
\emph{Theoretical Population Biology}
70 (2006), 237--243.

\bibitem[Shi]{Shi}
K. Shimatani,
On the measurement of species diversity incorporating species differences,
\emph{Oikos} 93 (2001), 135--147.

\bibitem[SP]{SP}
A.R. Solow, S. Polasky,
Measuring biological diversity,
\emph{Environmental and Ecological Statistics} 1 (1994), 95--107.

\bibitem[Tsa]{Tsa}
C. Tsallis,
Possible generalization of Boltzmann--Gibbs statistics,
\emph{Journal of Statistical Physics} 52 (1988), 479--487.

\bibitem[VN]{VN}
R.S. Varga, R. Nabben,
On symmetric ultrametric matrices,
in L. Reichel, A. Ruttan, R.S. Varga (eds.),
\emph{Numerical Linear Algebra},
pages 193--199,
Walter de Gruyter, 1993.

\end{thebibliography}
\end{document}